\documentclass[11pt]{article}
\usepackage{latexsym,amsmath,amscd,amssymb,graphicx,amsbsy,color,xcolor,amsfonts,amsthm}
\newtheorem{defn}{Definition}

\newtheorem{exam}{Example}
\newtheorem{claim}{Claim}
\begin{document}
\title{The wave-function description of the electromagnetic field\thanks{Supported in part by the German-Israel Foundation for Scientific Research and Development: GIF No. 1078-107.14/2009}}

\author{Yaakov Friedman \\Jerusalem College of Technology\\
Departments of Mathematics and Physics \\
P.O.B. 16031 Jerusalem 91160, Israel\\
e-mail: friedman@jct.ac.il}
\maketitle

\begin{abstract}

For an arbitrary electromagnetic field, we define a prepotential $S$, which is a complex-valued function of spacetime. The prepotential is a modification of the two scalar potential functions introduced by E. T. Whittaker. The prepotential is Lorentz covariant under a spin half representation. For a moving charge and any observer, we obtain a complex dimensionless scalar. The prepotential is a function of this dimensionless scalar. The prepotential $S$ of an arbitrary electromagnetic field is described as an integral over the charges generating the field. The Faraday vector at each point may be derived from $S$ by a convolution of the differential operator with the alpha matrices of Dirac. Some explicit examples will be calculated.  We also present the Maxwell equations for the prepotential.
\end{abstract}

\section{Introduction and motivation}\label{Intro}

$\;\;\;$

In general, the electromagnetic  field
tensor ${F}$, expressed by a $4\times 4$  antisymmetric matrix, is used to describe
the electromagnetic field intensity. This description involves 6 parameters.  An alternative way to describe an
electromagnetic  field is by use of the 4-potential. In a chosen gauge, the 4-potential
transforms as a 4-vector. The electromagnetic field tensor is then recovered by differentiating the 4-potential.
Choice of a gauge can reduce the degrees of freedom of this description to from 4 to 3.

In 1904 E. T. Whittaker introduced \cite{Whittaker} two scalar potential functions. Thus, he was able to reduce
the degrees of freedom of the electromagnetic  field description to 2.
He showed that the electromagnetic field can be expressed in terms of the
second derivatives of these functions. However, he was not able to find the covariance of his scalar potentials.
H. S. Ruse \cite{Ruse} improved  the result of Whittaker. In 2009 Y. Friedman  and S. Gwertzman  showed  \cite{FGW} that it is possible to combine these two scalar potential functions into one complex-valued function ${S}(x)$ on Minkowski space, which we call the \textit{prepotential of the electromagnetic field}. Moreover, they showed that this prepotential is invariant under a certain spin half representation of the Lorentz group. Thus, this prepotential provides a covariant description of an electromagnetic field with minimal degrees of freedom.

In 1953, influenced by M. Born, who was looking for the connection of the wave-function description of elementary particles in quantum mechanics and the electromagnetic field they generate, H. S. Green and E. Wolf introduced \cite{GreenWolf53} a complex scalar potential for an electromagnetic field.  They described the similarity of the expressions for energy and momentum densities between their potential and the wave function. They were unable, however, to find the connection between their potential and the Whittaker potentials.

An electromagnetic field is generated by a collection of moving charges.  Thus, a description of an electromagnetic field can be obtained by integrating the fields of moving charges. For a moving charge and observer at
point $x$, the position of the charge at the retarded time relative to the observer is a null-vector in spacetime.  We show that for each such vector, there is a complex dimensionless scalar which is invariant under a certain representation of the Lorentz group. The prepotential $S(x)$ is defined to be the logarithm of this scalar.

The Aharonov-Bohm effect indicates that there is a multiple-valued prepotential of the 4-potential of an electromagnetic field.
 In their paper \cite{AharonovBohm59}, a scalar function $S$  such that $\nabla S=(e\hbar/c)A$, where $A$ is the 4-potential of the electromagnetic field,  was introduced.
It has been shown that if  $\psi_0$ is the solution of the Schr\"odinger equation in the absence of an electromagnetic field, then the function $\psi=\psi_0e^{-iS/\hbar}$ is the solution of the equation in the presence
of the field, at least in a simply connected region in which the electromagnetic field vanishes.
In the magnetic Aharonov-Bohm experiment, however, the region outside the solenoid is not simply connected. This leads to a multi-valued $\psi$ in this experiment. As it was shown in \cite{FOst}, our prepotential is of the same type, but is defined for any electromagnetic field. Our prepotential is also multi-valued.

In classical mechanics, the negative of the gradient  of a scalar potential equals the force.
 This is true for forces which generate linear acceleration. Such forces are defined by a one-form
(since their line integral gives the work). The derivative of a scalar function
is also a one-form. Hence, the derivative of a potential can equal the negative of the
force. But in classical mechanics we also have rotating forces, which are described by
two-forms. Such forces cannot be expressed as derivatives of a scalar potential. For example, the electromagnetic field is not conservative in general. It generates both linear and rotational acceleration, and the electromagnetic tensor is a two-form. Hence, it is natural to assume that a kind of second derivative of a scalar potential  will define
the force tensor. Note that the usual differential of a gradient of a real-valued function is zero.
Therefore, the prepotential must be complex-valued, and we will need to define a Lorentz invariant conjugation of the gradient of the prepotential in order to obtain the 4-potential of the field.

Another important property of a prepotential of an electromagnetic field is its \textit{locality}. Note that the electromagnetic tensor $F_{\mu\nu}$ of a field of a moving charge depends on the position, velocity and acceleration, while the 4-potential $A_\mu$ depends only on the position and velocity of the source. Our prepotential $S$ depends only on the position of the source.

In section 2, we obtain a Lorentz group representation based on the complex electromagnetic field tensor. We will show that the regular representation $\pi$ of the Lorentz group can be decomposed as a product of two commuting representations $\tilde{\pi}$ and  $\tilde{\pi}^*$. In section 3, we will introduce the prepotential of an electromagnetic field and show its geometric meaning. We will also show the covariance of this prepotential under the action of the representation $\tilde{\pi}$. In section 4, we will study the connection between the prepotential and the Faraday vector of a field.
We will find the gauge of the prepotential. Explicit solutions for the prepotential and field of a charge at rest and a charged infinite rod will be found. Using computer algebra, we will derive the field of an arbitrary moving charge from its prepotential. The Maxwell equations for the prepotential will be derived in section 5 and we will demonstrate their use for deriving the prepotential and the field of current in straight wire. In section 6, we will show that the representation $\tilde{\pi}$ coincides with the spin half representation of the Lorentz group on the spinors. We also show that the matrices occurring in the description of the connection of the prepotential to the field are a representation of the Dirac $\alpha$-matrices.

\section{Lorentz group representation based on complex electromagnetic  field tensor}\label{secLor}

In flat Minkowski space $M$, the spacetime coordinates of an event are denoted by
$x^{\mu}\;(\mu=0,1,2,3)$, with $x^0=ct$. The Minkowski inner product is
$x\cdot y=\eta_{\mu\nu}x^{\mu}y^{\nu}$, where
the Minkowski metric is $\eta_{\mu\nu}=\operatorname{diag}(1,-1,-1,-1)$.
The usual Lorentz group representation  $\pi$ can be associated with an electromagnetic  field tensor $F_\alpha^\beta(\mathbf{E},\mathbf{B})$
\begin{equation}\label{EMMixedTensor}
F=F_\mu^\nu = \left(\begin{array}{cccc}
0 & E_1 & E_2 & E_3 \\
E_1 & 0 & cB_3 & -cB_2 \\
E_2 & -cB_3 & 0 & cB_1 \\
E_3 & cB_2 & -cB_1 & 0 \\
\end{array}\right),
\end{equation}
 where $\mathbf{E}$ denote the electric field intensity
 and  $\mathbf{B}$ the magnetic field intensity, as follows.

Since a magnetic field generates a rotation,
a generator of a rotation about the direction $\mathbf{n}\in \mathbb{R}^3$ is defined by  $F_\alpha^\beta(0,\mathbf{B})$, with $c\mathbf{B}=-\mathbf{n}$. Thus, the representation of the rotation
about the direction $\mathbf{n}$ with angle $\omega$ is given by the operator $\exp(F_\alpha^\beta(0,-\mathbf{n})\omega).$ For example, a
rotation $\mathfrak{R}^1$ about the $x^1$-axis (rotation in the $x^2x^3$ plane) is represented as
\begin{equation}\label{RepPiR1}
\pi(\mathfrak{R}^1)(\omega)=\Lambda_{23}(\omega)=\exp (\left(
                          \begin{array}{cccc}
                            0 & 0 & 0 & 0 \\
                            0 & 0 & 0 & 0 \\
                            0 & 0 & 0 & -1 \\
                            0& 0 & 1 & 0 \\
                          \end{array}
                        \right)\omega)=\left(
                                       \begin{array}{cccc}
                                         1 & 0 & 0 & 0 \\
                                        0 &  1 & 0 & 0 \\
                                         0 & 0 & \cos\omega &  -\sin\omega  \\
                                         0 & 0 &  \sin\omega  &  \cos\omega  \\
                                       \end{array}
                                     \right).
\end{equation}
Similarly, since an electric field generate boosts, a generator of a boost in the direction $\mathbf{n}$ can be identified with $F_\alpha^\beta(\mathbf{n},0)$. Thus, the representation of the boosts in the direction $\mathbf{n}$ with rapidity $\omega$ is given by the operator $\exp(F_\alpha^\beta(\mathbf{n},0 )\omega).$ For example,
boosts $\mathfrak{B}^1$ in the $x$-axis are represented as
\begin{equation}\label{RepPiB1}
\pi(\mathfrak{B}^1)(\omega)=\Lambda_{01}(\omega)=\exp (\left(
                          \begin{array}{cccc}
                            0 & 1 & 0 & 0 \\
                            1 & 0 & 0 & 0 \\
                            0 & 0 & 0 & 0 \\
                            0& 0 & 0 & 0 \\
                          \end{array}
                        \right)\omega)=\left(
                                       \begin{array}{cccc}
                                         \cosh\omega &  \sinh\omega & 0 & 0 \\
                                        \sinh\omega  &  \cosh\omega  & 0 & 0 \\
                                         0 & 0 & 1&0  \\
                                         0 & 0 & 0&1  \\
                                       \end{array}
                                     \right).
\end{equation}

Our complex prepotential $S(x)$ is a function $M \rightarrow \mathbb{C}$ on Minkowski space $M$.
Its gradient at any spacetime point  belongs to the complexified cotangent space at that
point, which we identify with $\mathbb{C}^4$.  Denote by $M_c$ the complex space $\mathbb{C}^4$
endowed with the bilinear complex-valued form $x\cdot y=\eta_{\mu\nu} x^\mu y^\nu$, which can
be considered as a complexification of Minkowski space. The bilinear form on $M_c$ is an extension of the Minkowski inner product.

Obviously, the representation $\pi$ of the Lorentz group on $M$ can be extended linearly to
a representation on $M_c$. We claim that this representation can be decomposed as follows.
\begin{claim}\label{pi-decomposition}
The representation $\pi$ of the Lorentz group on $M_c$ can be decomposed  into a product
\begin{equation}\label{DecLorents}
    \pi=\tilde{\pi}\tilde{\pi}^*
\end{equation}
of two commuting representations $\tilde{\pi}$ and $\tilde{\pi}^*$ on $M_c$.
\end{claim}
In order to prove this claim, we will decompose the generator $F^\alpha_\beta $ of the representation $\pi$ into a sum
\begin{equation}\label{DecGenLorents}
    F^\alpha_\beta=\frac{1}{2}\mathcal{F}^\alpha_\beta +\frac{1}{2}\bar{\mathcal{F}}^\alpha_\beta ,
\end{equation}
where  $\mathcal{F}^\alpha_\beta$ is  the complex electromagnetic tensor and its complex adjoint $\bar{\mathcal{F}}^\alpha_\beta$, defined below.

An electromagnetic field can be defined by the \textit{Faraday vector} $\mathbf{F}=\mathbf{E}+ic\mathbf{B}$. Note that since $i$ is a pseudo-scalar and $\mathbf{B}$
is a pseudo-vector, the expression $i\mathbf{B}$ is a vector which is independent of
the chosen  orientation of the space. The Faraday vector is used
to describe the Lorentz invariant field constants, see for example \cite{Landau}.
In \cite{FD}, Friedman and Danziger introduced a \textit{complex electromagnetic tensor} $\mathcal{F}^\alpha_\beta$  for the
description of an electromagnetic field, similar to the one introduced by Silberstein in \cite{Silberstein1} and
\cite{Silberstein2}. Complexified electromagnetic fields were also studied by A. Gersten \cite{Gerst} and played an important role in obtaining explicit solutions in \cite{FS} and \cite{F04} for motion of a charge in constant electromagnetic field.

Let $F^\alpha_\beta (\mathbf{E},\mathbf{B})$ be the usual electromagnetic tensor. The
complex electromagnetic tensor $\mathcal{F}^\alpha_\beta(\mathbf{E},\mathbf{B})$ is defined by $\mathcal{F}^\alpha_\beta(\mathbf{E},\mathbf{B})=F^\alpha_\beta (\mathbf{F},-i\mathbf{F})$
or
\begin{equation}\label{complex tensor}
\mathcal{F}^\alpha_\beta=\left(
                          \begin{array}{cccc}
                            0 & F_1 & F_2& F_3 \\
                            F_1 & 0 & -iF_3 & iF_2 \\
                            F_2& iF_3 & 0 & -iF_1 \\
                            F_3& -iF_2 & iF_1 & 0 \\
                          \end{array}
                        \right)=\sum_{j=1}^3 (\rho^j)^\alpha_\beta F_j\,,
\end{equation}
where $(\rho_j)^\alpha_\beta$ are the Majorana-Oppenheimer matrices (see \cite{Dvoeglasov})
   \[
(\rho^1)^\alpha_\beta=\left(
\begin{array}{cccc}
0 & 1 & 0 & 0 \\
1 & 0 & 0 & 0 \\
0 & 0 & 0 & -i \\
0 & 0 & i & 0 \\
\end{array}\right),\;
(\rho^2)^\alpha_\beta=\left(
\begin{array}{cccc}
0 & 0 & 1 & 0 \\
0 & 0 & 0 & i \\
1 & 0 & 0 & 0 \\
0 & -i & 0 & 0 \\
\end{array}\right),\]

\begin{equation}\label{rho3}
 (\rho^3)^\alpha_\beta=\left(
\begin{array}{cccc}
0 & 0 & 0 & 1 \\
0 & 0 & -i & 0 \\
0 & i & 0 & 0 \\
1 & 0 & 0 & 0 \\
\end{array}\right).
\end{equation}

Note that the electromagnetic tensor $F^\alpha_\beta$ is the real part of the complex tensor $\mathcal{F}^\alpha_\beta$. Equation (\ref{DecGenLorents}) holds now if we define the complex conjugate tensor to be $\bar{\mathcal{F}}^\alpha_\beta=\sum_{j=1}^3 (\bar{\rho}_j)^\alpha_\beta \bar{F}_j$.

The Majorana-Oppenheimer matrices can be derived from the connection of the 4-potential and the Faraday vector, as follows. It is known that if $A_\mu$ is the 4-potential of the electromagnetic field, then $E_j=F_{0j}=\partial_jA_0-\partial_0A_j= A_{0,j}-A_{j,0}$, where $\partial_\mu=\frac{\partial}{\partial x^\mu}$  and $B^j=(\nabla\times\mathbf{A})^j$, where $\nabla\times$ is the curl which is applied to the vector potential. Thus, the the $j$ component of the Faraday vector is connected to the potential as
\begin{equation}\label{FaradayAnd4potent}
 F^j= ( {\rho}^j)^{\alpha\beta}\partial_\alpha A_\beta=( {\rho}^j)^{\alpha\beta} A_{\beta,\alpha} \,,
\end{equation}
showing that the matrices $\rho^j$ connect the 4-potential to the Faraday vector of the electromagnetic field.
Using (\ref{complex tensor}), we can define a differential operator which connects the 4-potential with the the complex electromagnetic tensor $\mathcal{F}^\alpha_\beta$:
\begin{equation}\label{AtoCalF}
    \mathcal{F}^\alpha_\beta=\sum_{j=1}^3 (\rho^j)^\alpha_\beta ( {\rho}^j)^{\nu\mu}\partial_\nu A_\mu=\nabla\times A\,,
\end{equation}
where the \textit{curl} $\nabla\times$ on $M_c$ is defined as
\begin{equation}\label{curlDef}
   \nabla\times=\sum_{j=1}^3 (\rho_j)^\alpha_\beta ( {\rho}^j)^{\nu\mu}\partial_\nu\,.
\end{equation}

We will now derive a Lorentz group representation on $M_c$ based on the Majorana-Oppenheimer matrices. In Section 6 we will show that this representation can be identified with the known spin-half representation of the Lorentz group. To define this representation, it is enough to define the representation on the generators of boosts and rotations.

\begin{defn}\label{pi-tildeDefn}
Define a representation $\tilde{\pi}$ on $M_c$ by defining generators $\tilde{B}^j$ of boosts $\mathfrak{B}^j$ in the direction of $x^j$  to be $\frac{1}{2}\rho^j$ and generators $\tilde{R}^j$ of rotations $\mathfrak{R}^j$ about the direction $x^j$ to be $\frac{i}{2}\rho^j$ .
\end{defn}
\begin{claim}\label{pi-tildeRepr}
The representation $\tilde{\pi}$ is a Lorentz group representation  on $M_c$.
\end{claim}
Direct calculation shows that the operators $\tilde{R}^j=\frac{i}{2}\rho^j$ obey the same commutator relations as the generators of the rotation group $SO(3)$:
\[[\tilde{R}^j,\tilde{R}^k]=-\varepsilon^{jk}_l\tilde{R}^l,\]
where $\varepsilon^{ijk}$ is the Levi-Civita tensor.
Moreover, the $\tilde{B}^j=\frac{1}{2}\rho^j$ matrices obey the same commutator relations as the generators of boosts in the Lorentz group:
   \[ [\tilde{B}^j,\tilde{B}^k]=\varepsilon^{jk}_l\tilde{R}^l,\;\;\;\;
   [\tilde{R}^j,\tilde{B}^k]=\varepsilon^{jk}_l\tilde{B}^l\]
   This proves Claim \ref{pi-tildeRepr}.

The complex adjoints $\bar{\sigma}^j=\frac{-i}{2}\bar{\rho}^j$ and $\frac{1}{2}\bar{\rho}^j$ satisfy similar relations. This leads to a second Lorentz group representation on $M_c$.
\begin{defn}\label{pi-tildeDefnStar}
Define a representation $\tilde{\pi}^*$ on $M_c$ by defining generators $\tilde{B}^j$ of boosts $\mathfrak{B}^j$ in the direction of $x^j$  to be $\frac{1}{2}\bar{\rho}^j$ and generators $\tilde{R}^j$ of rotations $\mathfrak{R}^j$ about the direction $x^j$ to be $\frac{i}{2}\bar{\rho}^j$.
\end{defn}
Moreover, the two sets of operators $\{\rho^j\}$ and $\{\bar{\rho}^k\}$ commute:
\begin{equation}\label{CommutinfMO}
   [\rho^j,\bar{\rho}^k]=0.
\end{equation}
Thus, the representations  $\tilde{\pi}$ and $\tilde{\pi}^*$ commute.
 Since \[\exp F^\alpha_\beta =\exp(\frac{1}{2}\mathcal{F}^\alpha_\beta +\frac{1}{2}\bar{\mathcal{F}}^\alpha_\beta)=\exp(\frac{1}{2}\mathcal{F}^\alpha_\beta) \exp(\frac{1}{2}\bar{\mathcal{F}}^\alpha_\beta),\]
  the representation $\pi$, associated with the tensor $F^\alpha_\beta$, can be decomposed as a product (\ref{DecLorents}) of representations $\tilde{\pi}$ and $\tilde{\pi}^*$. This proves Claim \ref{pi-decomposition}.

In addition to the above commutation relations, Majorana-Oppenheimer matrices also satisfy \textit{anti-commutation relations}, which are very helpful for the calculation of the exponents of these matrices. The anti-commutation relations are
\begin{equation}\label{CAR}
  \{\rho^j ,\rho^k\}=\frac{1}{2}(\rho^j\rho^k+\rho^k \rho^j)=\delta^{jk}I\,,
\end{equation}
which implies that $(\rho^j)^2=I$. Using these relation and the power series expansion of the exponent function, the exponents of these matrices are
\[\exp (\rho^j\frac{\omega}{2})= \cosh \frac{\omega}{2} I +\rho^j\sinh \frac{\omega}{2}\]
\[\exp (\rho^j\frac{i\omega}{2})= \cos \frac{\omega}{2} I +i\rho^j\sin \frac{\omega}{2}\,.\]

We can define now explicitly the  representation under $\tilde{\pi}$ of the rotations and the boosts of the Lorentz group. For example, the rotation $\mathfrak{R}^1$ about the $x^1$-axis  or rotation in the $x^2x^3$ plane by an angle $\omega$ is represented by
\begin{equation}\label{RotTildePi}
 \tilde{\pi}(\mathfrak{R}^1)(\omega)=\Upsilon_{23}(\omega)=\exp (i\rho^1\frac{\omega}{2})= \cos \frac{\omega}{2} I -i\rho^1\sin\frac{\omega}{2}
\end{equation}
\[=\left(
                                       \begin{array}{cccc}
                                         \cos \omega/2 & -i\sin \omega/2 & 0 & 0 \\
                                        -i\sin \omega/2  &  \cos \omega/2 & 0 & 0 \\
                                         0 & 0 &  \cos \omega/2 & -\sin \omega/2 \\
                                        0&0& \sin \omega/2  &  \cos \omega/2\\
                                       \end{array}
                                     \right)\,.\]
The representation $\tilde{\pi}^*$ of this rotation is given by $\bar{\Upsilon}_{23}(\omega)$, the complex conjugate of the above matrix. In the $x^2x^3$ plane, both $\tilde{\pi}(\mathfrak{R}^1)$ and $\tilde{\pi}^*(\mathfrak{R}^1)$ define a rotation by an angle $\omega/2$, and their product $\Upsilon_{23}(\omega)\bar{\Upsilon}_{23}(\omega)=\Lambda_{23}(\omega)$, defined by (\ref{RepPiR1}), is a rotation by an angle $\omega$ in this plane. However, in the plane $x^0x^1$, the rotation $\Lambda_{23}(\omega)$ under $\pi$ is the identity, while both $\Upsilon_{23}(\omega)$ and $\bar{\Upsilon}_{23}(\omega)$ define a ``complex rotation" in the plane $\{x^0+ix^1:\;\;x^0,x^1\in \mathbb{R}\}$ by angles $-\omega/2$ and $\omega/2$, respectively, indicating the there is a rotation around the $x^1$ axis. This observation needs further understanding.

Similarly, the boost in the $x^1$-direction with parameter $\omega$ is
\begin{equation}\label{Boosttildepi}
 \tilde{\pi}(\mathfrak{B}^1)(\omega)=\Upsilon_{01}(\omega)=\exp (\rho^1\frac{\omega}{2})= \cosh \frac{\omega}{2} I +\rho^1\sinh \frac{\omega}{2}\end{equation}
\[=\left(
                                       \begin{array}{cccc}
                                         \cosh \omega/2 & \sinh \omega/2 & 0 & 0 \\
                                        \sinh \omega/2  &  \cosh \omega/2 & 0 & 0 \\
                                         0 & 0 &  \cosh \omega/2 &-i \sinh \omega/2 \\
                                        0&0& i\sinh \omega/2  &  \cosh \omega/2\\
                                       \end{array}
                                     \right).\]
 The representation $\tilde{\pi}^*$ of this boost is given by $\bar{\Upsilon}_{01}(\omega)$, the complex conjugate of the above matrix. In the $x^0x^1$ plane, both $\tilde{\pi}(\mathfrak{B}^1)$ and $\tilde{\pi}^*(\mathfrak{B}^1)$ define boosts in the $x^1$ direction with rapidity $\omega/2$, and their product $\Upsilon_{01}(\omega)\bar{\Upsilon}_{01}(\omega)=\Lambda_{01}(\omega)$, defined by (\ref{RepPiB1}), is a boost with rapidity $\omega$ in the $x^1$ direction. However, in the $x^2x^3$ plane, the boost $\Lambda_{01}(\omega)$ under $\pi$ is the identity, while both $\Upsilon_{01}(\omega)$ and $\bar{\Upsilon}_{01}(\omega)$ define ``complex boosts" in the plane $\{x^2+ix^3:\;\;x^2,x^3\in \mathbb{R}\}$, with rapidity  $-\omega/2$ and $\omega/2$, respectively, and the product of these boosts is the identity.

Note that the operators $\Upsilon_{\alpha\beta}$ are isometries on the space $M_c$. The representation  $\tilde{\pi}$, defined by the operators $\Upsilon_{\alpha\beta}$, is a spin 1/2 representation of the Lorentz group on $M_c$. Similarly, the representation $\tilde{\pi}^*$ is defined by the operators $\bar{\Upsilon}_{\alpha\beta}$, which are obtained by taking the complex conjugate of $\Upsilon_{\alpha\beta}$. This is also a spin 1/2 representation of the Lorentz group on $M_c$, and it commutes with the representation $\tilde{\pi}$. In section \ref{SecSpinors}, we will show that the representation $\tilde{\pi}$ can be identified with the usual representation of the Lorentz group on the \textit{spinors}.

Let  $F_\alpha ^\beta$ be the electromagnetic tensor of some electromagnetic field. This tensor is covariant under the representation $\pi$ of the Lorentz group. What is the covariance of the complex tensor $\mathcal{F}_\alpha ^\beta$ associated with $F_\alpha ^\beta$? We will show the following.
\begin{claim}\label{covarftildeF}
The covariance of the electromagnetic tensor $F_\alpha ^\beta$ under the representation $\pi$ is equivalent to the covariance of the corresponding tensor $\mathcal{F}_\alpha ^\beta$ under the representation $\tilde\pi$.
\end{claim}
Let $\Lambda$ be any element of the representation $\pi$ of the Lorentz group. By (\ref{DecLorents}), there is an element $\Upsilon$ of $\tilde{\pi}$ such that $\Lambda=\Upsilon\bar\Upsilon$. Using that $\mathcal{F},\Upsilon \in \operatorname{span} \rho^j$, these operators commute with any $\bar{\rho}^k$ and thus with $\mathcal{\bar F}, \bar\Upsilon$.  Thus,
\[ F' = \Lambda F\Lambda^{-1}= \frac{1}{2} (\Upsilon\bar\Upsilon\mathcal{F}\bar\Upsilon^{-1}\Upsilon^{-1} + \Upsilon\bar\Upsilon\mathcal{\bar F}\bar\Upsilon^{-1}\Upsilon^{-1})\]\[ = \frac{1}{2} (\Upsilon\mathcal{F}\Upsilon^{-1} + \bar\Upsilon\mathcal{\bar F}\bar\Upsilon^{-1}) = \frac{1}{2} (\mathcal{F'}+\mathcal{\bar F'}).\]
This proves Claim \ref{covarftildeF}.

\section{Definition of the prepotential}

\begin{figure}[h!]
  \centering
\scalebox{0.4}{\includegraphics{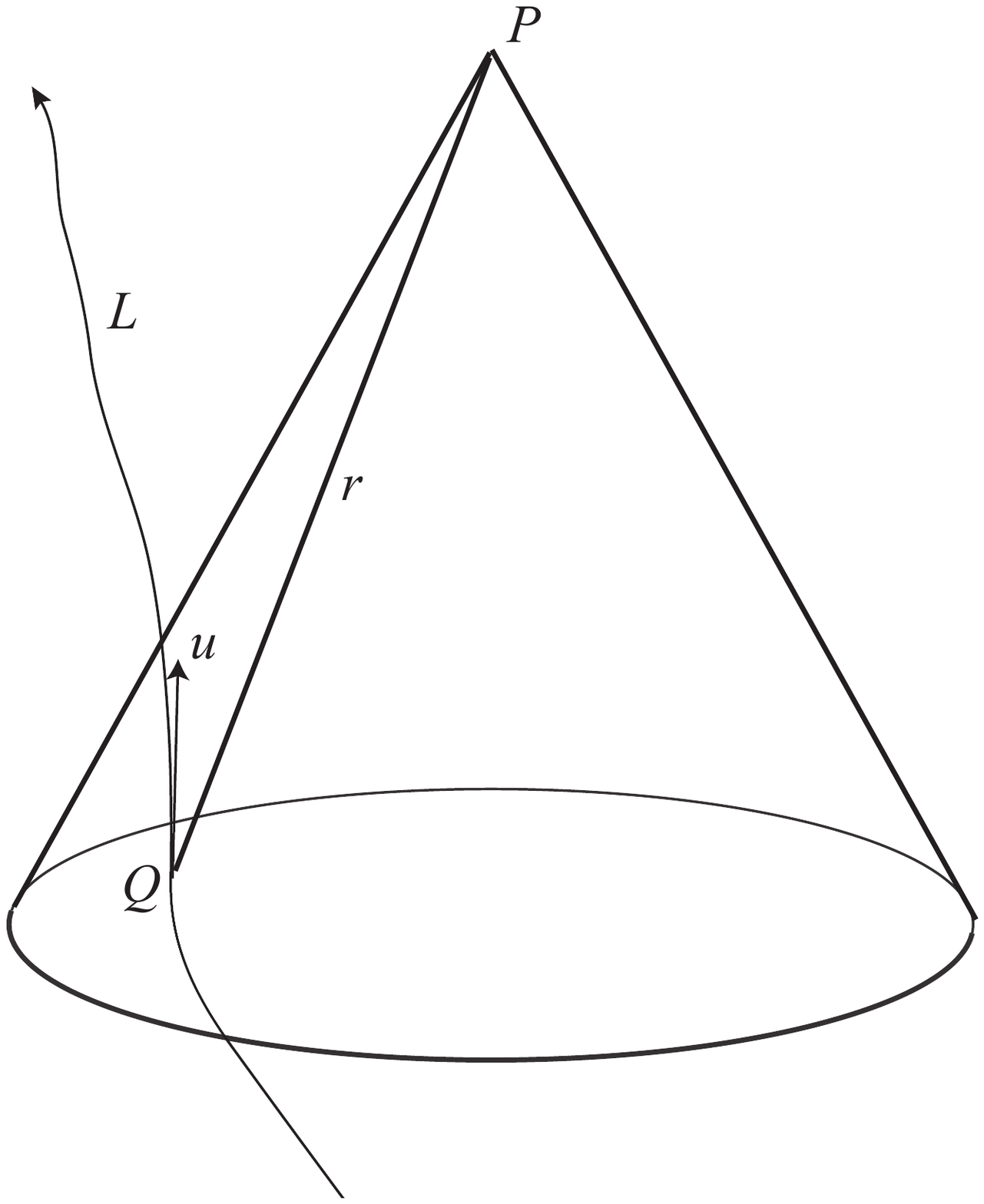}}
  \caption{The four-vectors associated with an observer and a moving charge.}\label{chargePotent}
\end{figure}

Denote by $P$ a point in Minkowski space $M$ at which we want to define
the prepotential. We will call $P$ the observer's point and denote his coordinates by $x=x^\mu$. Denote by $\check{x}(\tau)=\check{x}^\mu (\tau): \mathbb{R}\rightarrow M$ the worldline of the charge $q$ generating our electromagnetic field, as a function of its proper time. Let the point $Q=\check{x}(\tau(x))$ be the unique point of intersection of the past light cone at $P$ with the
worldline $\check{x}(\tau)$ of the charge. The time on the worldline of the charge corresponding to this intersection is uniquely determined by the point $x$. It is called the \textit{retarded time} and will be denoted by $\tau(x)$.  Note that radiation emitted at $Q$ at the retarded time will reach $P$ at time $t=x^0/c$ corresponding to this point. Our prepotential $S(x)$ will depend on the relative position
\begin{equation}\label{PositionDef}
   r(x)=x-\check{x}(\tau(x))
\end{equation}
of the charge at retarded time $\tau(x)$, see Figure \ref{chargePotent}.

 The relative position $r$ is a null vector. Hence,
  \begin{equation}\label{NullDecomp}
   (r^0+r^3)(r^0-r^3)=(r^1+ir^2)(r^1-ir^2).
\end{equation}
In this decomposition, we have split the four coordinates of $r$ into two groups: the first group contains the time component $r^0$ and the third spatial component $r^3$. The second group contains the first two spatial components $r^1$ and $r^2$.
  We can now define a dimensionless complex constant $\zeta$, as follows.
   \begin{defn}\label{zetaDef}
 For any null-vector $r$ in $M_c$, we define a dimensionless complex scalar $\zeta$ by
 \begin{equation}\label{zetaDefEq}
    \zeta(r)=\frac{r^0+r^3}{r^1-ir^2}=\frac{r^1+ir^2}{r^0-r^3}.
 \end{equation}
  For any $x\in M$ and any world-line $\check{x}(\tau)$, we define $\zeta (x)= \zeta(r(x))$, where the relative position $r(x)$ is defined by (\ref{PositionDef}).
 \end{defn}

  The scalar $\zeta(r)$ can be identified as a simple function of the stereographic projection of the direction of the vector part $\mathbf{r}$ of $r$ from the celestial sphere to the Argand plane, as follows. Since $r$ is a null vector, it can be decomposed as $(|\mathbf{r}|,\mathbf{r})$ and is defined uniquely by its vector part $\mathbf{r}$. Define $\hat{\mathbf{r}}=\mathbf{r}/|\mathbf{r}|$, a unit vector in the direction vector of $\mathbf{r}$, and a null vector $\hat{r}=(|\hat{\mathbf{r}}|,\hat{\mathbf{r}})$. Since $\zeta (r) $ is the same for any multiple of $r$, we have $\zeta (r)=\zeta (\hat{r})$. Then, as shown in \cite{PeR} p. 11, $\zeta (r)=\zeta (\hat{r})=e^{i\varphi} cot\frac{\theta}{2}$, where $\varphi,\theta$ are the angles of standard spherical coordinates of $\hat{r}$. This shows that $\zeta (r) $ depends only on the direction from the observer to the charge at the retarded time.

There is no way to associate to every null vector $r$ a non-trivial scalar which is invariant under the usual representation $\pi$. However, for the representation $\tilde{\pi}$ we have:
\begin{claim}\label{zetaInvar}
For any null vector $r$, the complex constant $\zeta(r)$, defined by (\ref{zetaDefEq}), is invariant under the representation $\tilde{\pi}$ on $M_c$.
\end{claim}
The boost $\Upsilon _{01}$ of $\tilde{\pi}$ in the $x^1$-direction is defined by (\ref{Boosttildepi}). Applying this boost to the vector $r$ yields
\[\Upsilon _{01} (r)=\cosh(\omega/2)(r^0,r^1,r^2,r^3)+\sinh(\omega/2)(r^1,r^0,-ir^3,ir^2).\]
Using (\ref{zetaDefEq}), we have
\[\Upsilon _{01} (\zeta)= \zeta(\Upsilon _{01} (r))=   \frac
{\cosh(\omega/2)(r^0+r^3)+\sinh(\omega/2)(r^1+ir^2)}{\cosh(\omega/2)(r^1-ir^2)+\sinh(\omega/2)(r^0-r^3)}=\]
\[\frac{r^0+r^3}{r^1-ir^2}\cdot
\frac{1+\tanh(\omega/2)(r^1+ir^2)/(r^0+r^3)}{1+\tanh(\omega/2)(r^0-r^3)/(r^1-ir^2)}
=\frac{r^0+r^3}{r^1-ir^2}\cdot
\frac{1+\tanh(\omega/2)/\bar{\zeta}}{1+\tanh(\omega/2)/\bar{\zeta}}
=\zeta,\]
showing that $\zeta$ is invariant under the boost $\Upsilon _{01}$. Similarly, one may show the
invariance of $\zeta$ under any boost and rotation of the representation $\tilde{\pi}$. This proves Claim \ref{zetaInvar}.

\begin{defn}\label{SdefDef} We define the \textit{prepotential} $S(x)$ at  point $x$ of an electromagnetic field generated by a moving charge $q$ as
\begin{equation}\label{Sdef}
   S(x)=\frac{q}{2}\ln \zeta (x)=\frac{q}{2}\ln \zeta(r (x)),
\end{equation}
where $\zeta (r)$ is defined by (\ref{zetaDefEq}) and the relative position of the charge $r(x)$ is defined by (\ref{PositionDef}).
\end{defn}
From Claim \ref{zetaInvar}, it follows that the prepotential is invariant under the representation $\tilde{\pi}$ on $M_c$.
Consider an electromagnetic field  generated by any charge distribution $\sigma (x)$. The scalar prepotential for this field is defined by
 \begin{equation}\label{scalPotGenera}
    S(x) =\frac{1}{2}\int\limits_{K^-(x)}\zeta(r)\sigma(x+r)d^3r,
\end{equation}
where $K^-(x)$ denotes the backward light-cone at $x$, and the integration is by the 3D volume on this cone.

The geometric meaning of this prepotential can be understood if we introduce \textit{relativistic bipolar coordinates} $(\varrho^0,\varrho^1,\theta,\varphi)$:
\[x^0=\varrho^0\cosh\theta,\;\;x^1=\varrho^1\cos \varphi,\;\;x^2=\varrho^1\sin \varphi,\;\;x^3=\varrho^0\sinh \theta.\]
The angle $\varphi$ in these coordinates is the same angle $\varphi$ occurring in polar and spherical coordinates. However, the angle $\theta$ differs from the angle $\theta$ of spherical coordinates. These coordinates fit better to our model as they give a simple form for the prepotential. The light-cone in these coordinates has the simple form $\varrho^0=\varrho^1=\varrho$, a hyperplane. For any null-vector $r$ on the light-cone, the invariant constant $\zeta(r)$, defined by (\ref{zetaDefEq}), in relativistic bipolar coordinates is
\[\zeta(r)=\frac{r^0+r^3}{r^1-ir^2}=\frac{\varrho e^{\theta}}{\varrho e^{-i\varphi}}=e^{\theta+i\varphi}.\]
The prepotential of a moving charge is
\[  S(x)=\frac{q}{2}(\theta+i\varphi),\] which is a multiple of the complex angle $\theta+i\varphi$ of the relative position $r(x)$ of the charge.

\section{ The prepotential and the Faraday vector of the field}

\subsection{Connection between the prepotential and the Faraday vector of a field}

The 4-potential will be defined as a conjugate of the gradient of the prepotential $S(x)$.
Since the prepotential $S(x)$ is invariant under the representation $\tilde{\pi}$, we want our
conjugation to commute with this representation. By Definition \ref{pi-tildeDefn} of the representation $\tilde{\pi}$, the conjugation needs to commute with any operator $\rho^j$. Using (\ref{CommutinfMO}), we can choose any operator $\bar{\rho}^k$ to define the conjugation. Since, in (\ref{NullDecomp}), and in the definition of $\zeta$, we choose the third spatial component to join the time component, also here we chose the $\tilde{\pi}$ \textit{covariant conjugation }$\mathcal{C}$ on $M_c$ to be given by multiplication by $\bar{\rho}^3$:
\begin{equation}\label{ConjOperDef}
  \mathcal{C}=
 (\bar{\rho}^3)^\alpha_\beta=\left(
\begin{array}{cccc}
0 & 0 & 0 & 1 \\
0 & 0 & i & 0 \\
0 & -i & 0 & 0 \\
1 & 0 & 0 & 0 \\
\end{array}\right)\,.
\end{equation}

The following diagram shows the derivation of the complex tensor $\mathcal{F}$ of the field from its prepotential $S(x)$:
 \[\begin{CD} S @>\nabla>> \nabla S @>\mathcal{C}>>A= \mathcal{C}(\nabla S) @>\nabla\times>> \mathcal{F}.
 \end{CD}\]
By differentiating the prepotential and taking its conjugate, we obtain the 4-potential $A$. Then we use (\ref{AtoCalF}) to get from $A$ to the complex tensor $\mathcal{F}$ of the field. The components $F_j$ of the Faraday vector can be found by (\ref{FaradayAnd4potent}) as
\begin{equation}\label{FfromS}
 F_j={\partial}_\nu (\rho^j)^{\mu\nu}(\bar{\rho}^3)_\mu^\lambda{\partial}_\lambda S\,.
\end{equation}
\begin{claim}\label{ClCovStoF}
The Faraday vector $ F_j$ defined by (\ref{FfromS}) is invariant under the representation $\tilde{\pi}$
\end{claim}
This claim follows from the following facts. Claim \ref{zetaInvar} implies that $S(x)$ is $\tilde{\pi}$ invariant. By definition of the conjugation, $(\bar{\rho}^3)_\mu^\lambda$ is  $\tilde{\pi}$  invariant. The remaining operators together form a  $\tilde{\pi}$ invariant operator.

The last expression can be simplified, if for each $j=1,2,3$, we introduce a new tensor $((\alpha^j)^{\mu\lambda}=(\rho^j)^{\mu\nu}(\bar{\rho}^3)_\mu^\lambda$. Then we can rewrite (\ref{FfromS}) as
\begin{equation}\label{FfromSAlf}
 F_j={(\alpha^j)^{\mu\lambda}\partial}_\nu {\partial}_\lambda S\,.
\end{equation}
The new tensors are
\[(\alpha^1)^{\mu\lambda}=\left(
\begin{array}{cccc}
0 & 0 & -i & 0 \\
0 & 0 & 0 & -1 \\
-i & 0 & 0 & 0 \\
0 & -1& 0 & 0 \\
\end{array}\right),\;\;(\alpha^2)^{\mu\lambda}=\left(
\begin{array}{cccc}
0 & i & 0 & 0 \\
i & 0 & 0 & 0\\
0 & 0 & 0 & -1  \\
0 & 0& -1& 0 \\
\end{array}\right)\]
\[(\alpha^3)^{\mu\lambda}=\left(
\begin{array}{cccc}
1 & 0 & 0 & 0 \\
0 & 1 & 0 & 0 \\
0 & 0 & 1 & 0 \\
0 & 0& 0 & -1 \\
\end{array}\right)\,.\]
Thus, (\ref{FfromSAlf}) defines the following explicit formulas for the connection between the prepotential $S(x)$ and the Faraday vector.
\begin{equation}
                               \begin{array}{l}
                              F_1=-2(S_{,13}+iS_{,02})  \\
                                F_2=-2(S_{,23}-iS_{,01}) \\
                               F_3= S_{,00}+S_{,11}+S_{,22}-S_{,33}\\
                             \end{array}\label{Fas derS}
\end{equation}

Based on these formulas, we obtain the following gauges for the prepotential.
\begin{claim}\label{claimGauge}
The prepotential $S(x)$ is not determined uniquely by the electromagnetic field. Let $g(x)$ be a function on $M_c$ from the following list, or a combination of such functions:
\begin{enumerate}
  \item $g(x)=g(x^1,x^2)$ is harmonic, that is, $g_{,11}+g_{,22}=0$,
  \item $g(x)=g(x^0,x^3)$ satisfies the wave equation, that is, $g_{,00}-g_{,33}=0$,
  \item $g(x)$ is of order 1 in $x^\mu$.
\end{enumerate}
Then the transformation of the prepotential
\begin{equation}\label{gagueTrans}
    S'(x)=S(x)+g(x)
\end{equation}
does not affect the field. A function $g(x)$ of this type will be called the \textit{prepotential gauge}.
\end{claim}
To prove the claim, it is enough to check that the field of the gauge $g(x)$ defined by (\ref{Fas derS}) is zero. In case $(i)$, we have $g_{,0}=g_{,3}=0$. Since each term of the components $F_1,F_2$ of the field in
(\ref{Fas derS}) involves differentiation by $x^0$ or by $x^3$, these components are zero. The third component $F_3$ is zero since $g_{,00}=g_{,33}=0$ and $g_{,11}+g_{,22}=0$. Similarly, for case $(ii)$, we have $g_{,1}=g_{,2}=0$. Since each term of the components $F_1,F_2$ of the field in
(\ref{Fas derS}) involve differentiation by $x^1$ or by $x^2$, these components are zero. The third component $F_3$ is zero since $g_{,11}=g_{,22}=0$ and $g_{,00}-g_{,33}=0$. Since each component of the field involves second derivatives, it will vanish for any function $g(x)$ of order 1 in $x^\mu$. This completes the proof of the claim.

We demonstrate now the use of  formulas  (\ref{Fas derS}) and Claim \ref{claimGauge} in calculating the field of a rest charge at the origin from its prepotential.
Consider a rest charge $q$ at the origin. The worldline of this charge is $\check{x}(\tau)=(\tau,0,0,0)$. Thus, the relative position of the charge is
 $r=(|x|,x^1,x^2,x^3)$, where $|x|=\sqrt{(x^1)^2+(x^2)^2+(x^3)^2}$.
From (\ref{zetaDefEq}) and (\ref{Sdef}),
\[S'(x)=\frac{q}{2}\ln\frac{|x|+x^3}{x^1-ix^2}=\frac{q}{2}\ln(|x|+x^3)-\frac{q}{2}\ln(x^1-ix^2)\,.\]
The function $\ln(x^1-ix^2)$ is harmonic, and, by Claim \ref{claimGauge}$(i)$, is a prepotential gauge.
\begin{exam} For a rest charge $q$ at the origin, we define the prepotential to be
\begin{equation}\label{ResTPrepot}
S(x)=\frac{q}{2}\ln(|x|+x^3)\,.
\end{equation}
\end{exam}
Note that in this case the prepotential is well defined for all $x\ne0$ and could be chosen to be real valued.
\begin{claim}\label{ClResPoint}
The prepotential $S(x)$ defined by (\ref{ResTPrepot}) defines the electromagnetic field of the rest charge $q$ at the origin to be $\mathbf{E}=\frac{\mathbf{r}}{|\mathbf{r}|^3}$ and $\mathbf{B}=0$. Moreover, $S(x)$ satisfies the wave equation $\square  S(x)=0$.
\end{claim}
Since $S(x)$ is independent of $x^0$, we have $S_{,0}=0$. Direct calculation shows that $\frac{\partial}{\partial x^j}|x|=\frac{x^j}{|x|}$. Thus, we get
\[S_{,1}=\frac{q}{2|x|}\frac{x^1}{|x|+x^3},\;\;\;S_{,2}=\frac{q}{|x|2}\frac{x^2}{|x|+x^3}\,,\]
and \[ S_{,3}=\frac{q}{2}\frac{1}{|x|+x^3}\left(\frac{x^3}{|x|}+1\right)=\frac{q}{2\vert x\vert}\,.\]
The 4-potential $A$ in this case is
\[A=\mathcal{C}\nabla S= \frac{q}{2\vert x\vert}(1, -i\frac{x^2}{|x|+x^3},i\frac{x^1}{|x|+x^3},0)\,.\]
This potential behaves similarly to the usual 4-potential: when $|x|$ approaches infinity, it behaves like $\frac{1}{|x|}$.

By use of (\ref{Fas derS}), we get
\[ F_1=-2(S_{,13}+iS_{,02})=-2\frac{\partial}{\partial x^1}S_{,3}=-2\frac{\partial}{\partial x^1}
\frac{q}{2|x|}=\frac{q x^1}{|x|^3}\,,\]
\[ F_2=-2(S_{,23}-iS_{,01})=-2\frac{\partial}{\partial x^2}S_{,3}=-2\frac{\partial}{\partial x^2}
\frac{q}{2|x|}=\frac{q x^2}{|x|^3}\,.\]

To calculate $F_3$ using (\ref{Fas derS}), we need first to calculate $S_{,\mu\mu}$.
\[S_{,11}=\frac{\partial}{\partial x^1}\frac{q}{2}
\frac{x^1}{|x|(|x|+x^3)}=\frac{q}{2}\frac{|x|^2+x^3|x|-2(x^1)^2-x^3(x^1)^2/|x|}{|x|^2(|x|+x^3)^2},\]
\[S_{,22}=\frac{\partial}{\partial x^2}\frac{q}{2}
\frac{x^2}{|x|(|x|+x^3)}=\frac{q}{2}\frac{|x|^2+x^3|x|-2(x^2)^2-x^3(x^2)^2/|x|}{|x|^2(|x|+x^3)^2}.\]
This implies that
\[S_{,11}+S_{,22}=\frac{q x^3}{2|x|^3}\,.\]
Since $S_{,00}=0$ and $S_{,33}=\frac{-q x^3}{2|x|^3}$, equation (\ref{Fas derS}) yields
\[F_3=\frac{q x^3}{|x|^3}\]
and $\square  S(x)=S_{,00}-(S_{,11}+S_{,22})-S_{,33}=0$. This proves the claim.

As a second example, we will find now the prepotential of an infinitely long charged rod. We start with a prepotential
of a finite rod on the $x^3$ axis, from $x^3=-L$ to $x^3=L$, with charge density $\sigma$. Denote by $l$ the position of the charge on the rod, so that $-L<l<L$. The worldline of the charge is $\check{x}(\tau)=(\tau, 0,0,l)$, implying that $r=(\xi,x^1,x^2,x^3-l)$, with
  \begin{equation}\label{xiDef}
 \xi(l)=\sqrt{(x^1)^2+(x^2)^2+(x^3-l)^2}=\sqrt{\varrho^2+(x^3-l)^2}\,,
\end{equation}
where $\varrho=\sqrt{(x^1)^2+(x^2)^2}.$ Thus, from (\ref{zetaDefEq}), the zeta factor for the charge is \[\zeta(r)=\frac{\xi(l)+x^3-l}{x^1-ix^2},\]
and the prepotential of the rod is
\begin{equation}\label{PrepRod1}
    S(x)=\frac{\lambda}{2}\int_{-L}^L \ln\frac{\xi(l)+x^3-l}{x^1-ix^2}dl=\frac{\lambda}{2}\int_{-L}^L \ln(\xi(l)+x^3-l)dl-\frac{\lambda}{2}\int_{-L}^L \ln(x^1-ix^2)dl\,,
\end{equation}
with $\xi$ defined by (\ref{xiDef}).

The second integral is equal to $2L \ln(x^1-ix^2)$, which is a harmonic function in $x^1,x^2$, and, by Claim \ref{claimGauge}$(i)$, is a gauge and will be removed from the prepotential. For the first integral, we use a substitution $z=x^3-l$, implying $dl=-dz$. After the substitution, the first integral is
  \[-\int_{x^3+L}^{x^3-L} \ln(\sqrt{\varrho^2+z^2}+z)dz=\sqrt{\varrho^2+z^2}-z\ln(\sqrt{\varrho^2+z^2}+z),\]
   and the prepotential (\ref{PrepRod1}) is equivalent to
\[
    S(x)=\frac{\lambda}{2}\left( (x^3+L)\ln(\xi(-L)+x^3+L)-(x^3-L)\ln(\xi(L)+x^3-L) +\xi(L)-\xi(-L)\right)\,.\]

Using that $\xi(L)+x^3-L=\frac{\varrho^2}{\xi(L)-x^3+L}$ and $\xi(L)-\xi(-L)=\frac{-4x^3L}{\xi(L)+\xi(-L)}$, we get
\[S(x)=\frac{\lambda}{2}\left(-x^3\ln\varrho^2+L \ln\varrho^2-\frac{4x^3L}{\xi(L)+\xi(-L)}\right)\]\[ +\frac{\lambda}{2}\left((x^3+L)\ln(\xi(-L)+x^3+L)+(x^3-L)\ln(\xi(L)-x^3+L)\right).\]
The expression $L\ln \varrho^2=L \ln((x^1)^2+(x^2))$ is a harmonic function in $x^1,x^2$, and, by Claim \ref{claimGauge}$(i)$, is a gauge and will be removed from the prepotential.

In order to be able to pass to the limit $L\rightarrow\infty$, we will subtract from the prepotential a gauge function with a behavior similar to $S(x)$ for large $L$. An example of such a function is $2x^3\ln 2L-2x^3$, which by Claim \ref{claimGauge}$(iii)$ is a gauge. We obtain
\[S(x)=\frac{\lambda}{2}\left(-x^3\ln\varrho^2-\{\frac{4x^3L}{\xi(L)+\xi(-L)}-2x^3\}\right)\]
\[ +\frac{\lambda}{2}\left((x^3+L)\ln(\xi(-L)+x^3+L)+(x^3-L)\ln(\xi(L)-x^3+L)-2x^3\ln 2L\right).\]
Since $\lim_{L\to\infty}\xi(\pm L)/L=1$, we have
\[\lim_{L\to\infty}\frac{4x^3L}{\xi(L)+\xi(-L)}-2x^3=\lim_{L\to\infty}\frac{4x^3}{\xi(L)/L+\xi(-L)/L}-2x^3=0\]
and
\[\lim_{L\to\infty}(x^3+L)\ln(\xi(-L)+x^3+L)+(x^3-L)\ln(\xi(L)-x^3+L)-2x^3\ln 2L=0\]
Thus, the prepotential of an infinite rod is
\begin{equation}\label{RodInfPrep}
   S(x)=-\frac{\lambda}{2}x^3\ln\varrho^2=-\lambda x^3\ln\sqrt{(x^1)^2+(x^2)^2}\,.
\end{equation}

By applying formulas (\ref{Fas derS}) to this prepotential, we get
\[
                               \begin{array}{l}
                              F_1=-2(S_{,13}+iS_{,02})=\lambda \frac{2x^1}{\varrho^2} \\
                                F_2=-2(S_{,23}-iS_{,01})=\lambda \frac{2x^2}{\varrho^2} \\
                               F_3= S_{,00}+S_{,11}+S_{,22}-S_{,33}=0\,,
                             \end{array}\]
which implies that the electric field is $\mathbf{E}(x)=\frac{2\lambda }{\varrho^2}(x^1,x^2,0)$ and that there is no magnetic field, as is known for the field of a charged rod.

\subsection{The  Faraday vector  $F$ from the prepotential of a moving charge}

Consider now any charge $q$ with worldline $\check{x}(\tau)$. In order to be able to calculate the field of this charge from its prepotential $S(x)$ defined by (\ref{Sdef}), we will use formulas (\ref{Fas derS}). To apply this formula, we have to calculate some second partial derivatives of  $S(x)$. The difficulty is that $S$ is a function of $\zeta(r(x))$, and $r(x)$ depends also on the worldline $\check{x}(\tau)$. The idea is to present these derivatives in covariant form. This means that we want to express them by covariant vectors, namely, the relative position $r$ of the observer with respect to the charge at the retarded time $\tau(x)$, the 4-velocity $u=\frac{d\check{x}}{d\tau}(\tau(x))$ and the 4-acceleration $a=\frac{du}{d\tau}(\tau(x))$ at the retarded time.

From the definition (\ref{Sdef}) of the prepotential $S(x)$ its derivative by $x^\mu$ is
\begin{equation}\label{PotDerivat1}
    S(x)_{,\mu}=\frac{q}{2}\frac{\zeta_{,\mu}}{\zeta}=\left( \frac{r^0_{,\mu} +  r^3_{,\mu}}{r^1 -ir^2} -\frac{(r^0+r^3)(r^1_{,\mu}-ir^2_{,\mu})}{(r^1 -ir^2)^2}\right)\frac{r^1 -ir^2}{r^0+r^3}
=\frac{ r^0_{,\mu} +  r^3_{,\mu}}{r^0+r^3} -\frac{r^1_{,\mu}-ir^2_{,\mu}}{r^1 -ir^2} \,.
\end{equation}
Thus, to be able to calculate the derivatives of  $S(x)$, we need a formula for the derivatives of the relative position $r(x)$ defined by (\ref{PositionDef}). The partial derivative $r_{,\mu}$ of the position vector $r(x)$ by $x^\mu$ is
\[r_{,\mu}=x_{,\mu}-y_{,\mu}(\tau(x))= x_{,\mu}-\partial y/\partial \tau \,\tau_{,\mu}= x_{,\mu}-u \tau_{,\mu}.\]
Since $r(x)$ is null, we have $r\cdot r_{,\mu}=0$, where $a\cdot b=\eta_{\mu\nu}a^\mu b^\nu$ denote the Minkowski inner product. Thus,
\[0=r\cdot r_{,\mu}=r\cdot x_{,\mu}- (r\cdot u) \tau_{,\mu},\]
which imply that the derivative of the retarded time is
\begin{equation}\label{tauDer}
   \tau(x)_{,\mu}=\frac{r_\mu}{r\cdot u}.
\end{equation}
This mean that varying the observers position by $dx^\mu$ will cause the parameter $\tau$ to change infinitesimally by $\frac{r_\mu}{r\cdot u}dx^\mu$. Since $(r)^2=0$ and $(u)^2=1$, from the Cauchy-Schwartz inequality it follows that $r\cdot u$ is always non-zero. So, equation (\ref{tauDer}) is always well defined.

Substituting this into the derivative of the position vector we get
\begin{equation}\label{PositionDer}
   r^\nu_{,\mu}=x^\nu_{,\mu}-\frac{r_\mu u^\nu}{r\cdot u}=\delta^\nu_\mu-\frac{r_\mu u^\nu}{r\cdot u},
\end{equation}
where $\delta^\nu_\mu$ denotes the Kronecker delta function. This formula could be understood as follows. An infinitesimal variation  $dx^\mu$ of the observers position causes also a change of the charge position at the retarded time. The relative position $r$ under such variation is effected by the two changes. The first term in (\ref{PositionDer}) the effect of infinitesimal variation of the observers position, while the second one express change of the charge position at the retarded time.

Substituting derivative of the position vector (\ref{PositionDer}) in (\ref{PotDerivat1}) we obtain a covariant form for the derivative of the prepotential $S(x)$ as a function of the the relative position $r$ and the 4-velocity of the charge $u$ at the retarded time.
To calculate the electromagnetic field of of a moving charge from its prepotential $S(x)$, according (\ref{Fas derS})
we need to calculate some second derivatives of $S(x)$. This could be done by differentiating $S_{,\mu}$. The expression of $S_{,\mu}$ involve involve only non-constant vectors $r$ and $u$. The derivatives of $r$ are given by (\ref{PositionDer}). The derivatives of $u$, by use of (\ref{tauDer}), are
\begin{equation}\label{uDer}
    u_{,\mu}=\frac{\partial u}{\partial \tau}\tau (x)_{,\mu}=\frac{a r_\mu}{r\cdot u}\,.
\end{equation}
Since an infinitesimal change of the observers position by $dx^\mu$ causes the change of $\tau$  by amount $\frac{r_\mu}{r\cdot u}dx^\mu$, the variation of the 4-velocity $u$ of the charge caused by this change is a multiple of the acceleration $a$ of the charge by this amount.

To simplify the result of the differentiation by use of computer algebra we can  introduce for any $j=1,2,3$  an anti-symmetric scalar product on vectors of $M_c$ by
\begin{equation}\label{antisysmpriductDef}
   [a,b]_j=(\rho^j)_{\alpha\beta}a^\alpha b^\beta=\rho^j a\cdot b,\;\;\;\;a,b\in M_c \,.
\end{equation}
By use of computer algebra (\ref{Fas derS}) and using the above formulas for the derivatives we get
\begin{equation}\label{FmovigChargeFronSfinal}
F^j=q\left(\frac{\left[r,u\right]_j}{(r\cdot u)^3}+\frac{\left[r,a\right]_j(r\cdot u)-\left[r,u\right]_j(r\cdot a)}{2(r\cdot u)^3}\right)
\end{equation}

Using $1+3$ decomposition of 4-vectors it can be shown that these formulas coincide with the know field of a moving charge as derived from the Lienard-Wiecher potential, see \cite{Jackson} p.664. Thus, we have shown that the prepotential $S(x)$ of a moving charge defined by (\ref{Sdef}) produces by use of (\ref{Fas derS}) the known electromagnetic field of such charge. Moreover, the superposition principle and that the connection of the prepotential to the field is linear, for a general electromagnetic field, which is produce by a collection of moving charges, by applying  (\ref{Fas derS}) to the prepotential $S(x)$ defined by formula (\ref{scalPotGenera}) will recover properly this field.
%
%

%

\section{The Maxwell equations of the prepotential}

In this section  we define the  Maxwell equations for the prepotential $S(x)$ of the electromagnetic field.
Consider an electromagnetic field generated by sources $J^\mu=(\rho,j^1,j^2,j^3)$.
The pair of the scalar Maxwell equations
\[\nabla\cdot \mathbf{E}=\rho,\;\;\nabla\cdot \mathbf{B}=0\]
for the Faraday vector $\mathbf{F}=\mathbf{E}+ic\mathbf{B}$ can be rewritten as one equation
\begin{equation}\label{Max scal}
    \nabla\cdot \mathbf{F}=\rho
\end{equation}
Substitute the values of $\mathbf{F}$ from (\ref{Fas derS})  to get
\[F_{1,1} +F_{2,2}+F_{3,3}=-2S_{,113}-2iS_{,102}-2S_{,223}+2iS_{,201}+S_{,300}+S_{,311}+S_{,322}-S_{,333}\]\[=
  S_{,003}- S_{,113}-S_{,223}-S_{,333} =\rho\,.\]
 So, the first Maxwell equation for $S(x)$ is:
\begin{equation}\label{Max S 0}
   \partial_3( \square  S)=\rho
\end{equation}

The vector Maxwell equations are
\begin{equation}\label{Max vec}
    \nabla\times \mathbf{E}+\dot{\mathbf{B}}=0,\;\;\nabla\times \mathbf{B}-\dot{\mathbf{E}}=\mathbf{j}.
\end{equation}
They can be rewritten by use of the Faraday vector as one equation
\begin{equation}\label{Max scal}
    \mathbf{F}_{,0}+i \nabla\times \mathbf{F}=-\mathbf{j}.
\end{equation}

Substitute the values of $\mathbf{F}$ from (\ref{Fas derS}))  for the first component to get
\[F_{1,0} +iF_{3,2}-iF_{2,3}=-2S_{,013}-2iS_{,002}+i(S_{,200}+S_{,211}+S_{,222}-S_{,233})+2i(S_{,323}-iS_{,301})\]\[=
   i(-S_{,002}+S_{,211}+S_{,222}+S_{,233})=-j^1\,.\]
 So, the second Maxwell equation for $S(x)$ is:
\begin{equation}\label{Max S 1}
    -i\partial _2( \square  S)=-j^1\,.
\end{equation}
 For the second component we get
\[F_{2,0} +iF_{1,3}-iF_{3,1}=-2S_{,023}+2iS_{,001}-2i(S_{,313}+iS_{,302}-i(S_{,100}+S_{,111}+S_{,122}-S_{,133})\]
\[= i(S_{,100}-S_{,111}-S_{,122}-S_{,133})=-j^2\,.\]
 So, the third Maxwell equation for $S(x)$ is:
\begin{equation}\label{Max S 2}
   i\partial _1( \square  S)=-j^2\,.
\end{equation}
 For the third component we get
\[F_{3,0} +iF_{2,1}-iF_{1,2}=S_{,000}+S_{,011}+S_{,022}-S_{,033}-2i(S_{,123}-iS_{,101})+2i(S_{,213}+iS_{,202})\]
\[=  S_{,000}-S_{,011}-S_{,022}-S_{,033}=-j^3\,.\]
 So, the forth Maxwell equation for $S(x)$ is:
\begin{equation}\label{Max S 3}
   \partial_0( \square  S)=-j^3\,.
\end{equation}

Note since the left-hand side of these equations is a co-vector we lower the index of the sources of the field by $J_\mu=(\rho,-j^1,-j^2,-j^3)$. We will also apply the conjugation $\mathcal{C}$ defined by (\ref{ConjOperDef}) to the gradient $\partial$ to get
\[ \mathcal{C}^\nu_\mu \partial_\nu =(\partial_3,-i\partial_2,i\partial_1,\partial_0)\,.\]
With this notation we can rewrite our 4 Maxwell equations (\ref{Max S 0}), (\ref{Max S 1}),(\ref{Max S 2}) and
(\ref{Max S 3}), as
\begin{equation}\label{Maxwell eqn}
  \mathcal{C}^\nu_\mu \partial_\nu ( \square  S(x))=J_\mu(x)\,,
\end{equation}
or in operator form
\begin{equation}\label{Maxwell eqnOper}
  \mathcal{C}\nabla( \square  S(x))=J(x)\,.
\end{equation}
Our Maxwell equations (\ref{Maxwell eqnOper})have some similarity to the ones obtained by use of Clifford algebras, see \cite{Baylis}.

The Maxwell equation(s) for the prepotential we obtained is a third order system of equations. But this system can be decomposed into two systems of equations. The first system is a first order one $\mathcal{C}\nabla \Phi=J$. Using that
$\mathcal{C}$ is a symmetry, meaning $\mathcal{C}^2=I$, this system can be rewritten as $\nabla \Phi(x)=\mathcal{C}J(x)$. The meaning of such system is looking for a scalar potential $\Phi(x)$ for a vector-field $\mathcal{C}J(x)$. A similar formula is obtained in \cite{Barut} p.98 for the spinor form of Maxwell equation describing the connection of the sources of the field to the Faraday vector. The second system is a  known second order system $\square  S(x)=\Phi(x)$ for $\Phi(x)$, solved in the previous step. Thus, equation (\ref{Maxwell eqnOper}) is equivalent to
\begin{equation}\label{MaxTwoEqua}
    \nabla \Phi(x)=\mathcal{C}J(x),\;\;\square  S(x)=\Phi(x)\,.
\end{equation}

Claim \ref{ClResPoint} show that $S(x)$ satisfy the wave equation $\square  S(x)=0$ outside the rest charge. By using the covariance of such property, this will hold for the prepotential of any moving charge.  Using the superposition principle the prepotential of any electromagnetic field will satisfy the wave equation
\begin{equation}\label{waveS}
    \square  S(x)=0
\end{equation}
for $x$ outside the sources of the field.

We will demonstrate now the use of the Maxwell equations (\ref{Maxwell eqnOper}) for derivation of the prepotential and the field of a current $I$ in a thin long straight wire. We will assume that the wire is positioned on the $x^3$-axis.  The field of such current is generated by negatively charged electrons with line density $\tilde{\lambda}$ moving with velocity $\mathbf{v}$ in the negative direction of the $x^3$-axis. These charges are compensated by positive charges of the same density resting in the wire. To express the fact that the wire is thin we will use 2D Dirac-delta $\delta(\varrho)$ for $\varrho=\sqrt{(x^1)^2+(x^2)^2}$ with the property that the integral of this function over a domain $D$ in $x^1,x^2$ plane is equal 1 if the point $(0,0)$ is in $D$ and 0 otherwise.

The current density of the electron sources of the field are given by
$ j(e)^\mu=-\tilde{\lambda}\delta(\varrho)\gamma(1,0,0,-v/c),$
where $\gamma=\gamma(v)=(1-v^2/c^2)^{-1/2}$. So, the effective density will be $\lambda=\tilde{\lambda}\gamma$. For the positive charges
$ j(p)^\mu=\lambda\delta(\varrho)(1,0,0,0).$
So, the total current of the sources of the field will be $ j^\mu=\lambda\delta(\varrho)(0,0,0, \frac{v}{c})$ and using that $I=\lambda v$,  $ j^\mu=\delta(\varrho)(0,0,0, \frac{I}{c}).$
The equation for $\Phi(x)$ from (\ref{MaxTwoEqua}) become
\[\Phi_{,0}=- \frac{I}{c}\delta(\varrho),\;\;\Phi_{,1}= \Phi_{,2}=\Phi_{,3}=0\,.\]
Integrating the first equation and using that the average value of the Dirac delta is zero we get $\Phi(x)=-\frac{I}{c}\delta(\varrho)x^0 .$

Now we need to solve the equation  (\ref{MaxTwoEqua}) $\square  S(x)=\Phi(x)$ for the prepotential $S(x)$. From \cite{Landau2} it follows that the electromagnetic field must be normal to the wire, implying $F_3=0$ and by use of (\ref{Fas derS}) that $ S_{,00}+S_{,11}+S_{,22}-S_{,33}=0$. Subtracting from the last  equation $\square  S(x)=\Phi(x)$ we get
\begin{equation}\label{prepEqWire1}
   S_{,11}+S_{,22}=-\frac{1}{2}\Phi(x)=\frac{I}{2c}\delta(\varrho)x^0
\end{equation}
and adding these equations we get
\begin{equation}\label{prepEqWire2}
   S_{,00}-S_{,33}=\frac{1}{2}\Phi(x)=-\frac{I}{2c}\delta(\varrho)x^0\,.
\end{equation}

We can look for a solution for  equations (\ref{prepEqWire1}) and (\ref{prepEqWire2}). We will decompose the prepotential $S(x)$ as a sum of two components $S(x)=S_1(x)+S_2(x)$, where each component will satisfy one of the equations and vanish by the differential operator of the second one. We will look  for $S_1(x)$  of the form $S_1(x)=\frac{I}{2c}x^0f(x^1,x^2)$ for some function $f(x^1,x^2)$. The equation (\ref{prepEqWire1}) for $f$  become $\nabla ^2f(x^1,x^2)=\delta(\varrho)$. It is known (\cite{arfken} p.760) that the solution of such equation is $f(x^1,x^2)=\frac{-1}{2\pi}\ln\varrho$. Thus, $S_1(x)=\frac{I}{4\pi c}x^0\ln\varrho$ satisfy equation (\ref{prepEqWire1}) and vanish by the differential operator of equation (\ref{prepEqWire2}).

If we take  $S_2(x)$ to be of the form $S_2(x)=\frac{I}{2c}\delta(\varrho)x^0g(x^3)$, then such function will vanish differential operator of (\ref{prepEqWire1}) and for (\ref{prepEqWire2}) will give $g(x^3)=-(x^3)^2/2$.
Thus the prepotential $S(x)$ for a straight wire is
\begin{equation}\label{PrepStWire}
   S(x)=\frac{I}{4c}x^0\left(\frac{1}{\pi}\ln\varrho- (x^3)^2\delta(\varrho)\right)\,.
\end{equation}

By used of (\ref{Fas derS}) we can now calculate the Faraday vector of the field outside the wire. Since $\ln\varrho$ is harmonic outside the wire we have $F_3= S_{,00}+S_{,11}+S_{,22}-S_{,33}=0.$ Direct calculation give
$ F_1=-2(S_{,13}+iS_{,02})=-i\frac{I}{2\pi c}\frac{x^2}{\varrho^2}$ and
$ F_2=-2(S_{,23}-iS_{,01})=i\frac{I}{2\pi c}\frac{x^1}{\varrho^2}.$
 Thus, the electric field is zero and  the magnetic field is
 \[B_1=\frac{ I}{2\pi c^2} \frac{-x^2}{\varrho^2},\;\; B_2=\frac{ I}{2\pi c^2} \frac{x^1}{\varrho^2},
 \;\; B_3=0.\]
 This corresponds to the known formulas for such field (see \cite{Feynman} p.14-4).

\section{Prepotential, spinors and the Dirac matrices}\label{SecSpinors}

The prepotential $S(x)$ of a moving charge that we introduced was based on the decomposition (\ref{NullDecomp}) of any null vector in $M_c$. This decomposition involve four terms, which could be considered as coordinated of the null vector $r$ in a new basis. This basis on  $M_c$ is called the \textit{ Bondi null tetrad} and is called also Newman-Penrose basis. It  is defined as
\[\mathbf{ n}_0=\mathbf{n}=\frac{1}{\sqrt{2}}(\mathbf{e}_0+\mathbf{e}_3),\;
    \mathbf{ n}_1=\bar{\mathbf{m}}=\frac{1}{\sqrt{2}}(\mathbf{e}_1-i\mathbf{e}_2),\;\]
\begin{equation}\label{NPbasisNotaion}
      \mathbf{ n}_2=\mathbf{m}=\frac{1}{\sqrt{2}}(\mathbf{e}_1+i\mathbf{e}_2),\;
      \mathbf{ n}_3=\mathbf{l}=\frac{1}{\sqrt{2}}(\mathbf{e}_0-\mathbf{e}_3).
\end{equation}
For the significance of the Bondi tetrad see \cite{PeR}, \cite{ODonnell} and
\cite{FriedmanNP}. All vectors $ \mathbf{ n}$ are null vectors in $M_c$. Note that application of complex conjugation on $M_c$, which is equivalent
to replacing $i$ with $-i$, maps the Bondi tetrad into itself, but exchanges $\mathbf{ n}_1$ with
$\mathbf{ n}_2$. This mean that also here $i$ is a pseudo-scalar which with the change
of orientation, that may be expressed by change of the order of the basis vectors, changes its sign.

We will denote  by $y^\mu$ the coordinates of a vector in $M$ with respect
to the Bondi tetrad, meaning $x^\mu \mathbf{e}_\mu=y^\mu \mathbf{n}_\mu.$
Then, the relation between the coordinates is
\begin{equation}\label{r in term q}
   x^0=\frac{1}{\sqrt{2}}(y^0+y^3), \; x^1=\frac{1}{\sqrt{2}}(y^1+y^2),\;
    x^2=i\frac{1}{\sqrt{2}}(y^2-y^1),\;
    x^3=\frac{1}{\sqrt{2}}(y^0-y^3),
\end{equation}
or inversely,
\begin{equation}\label{q in term r}
   y^0=\frac{1}{\sqrt{2}}(x^0+x^3), \;
   y^1=\frac{1}{\sqrt{2}}(x^1+ix^2),\; y^2=\frac{1}{\sqrt{2}}(x^1-ix^2),\;
   y^3=\frac{1}{\sqrt{2}}(x^0-x^3).
\end{equation}

The coordinate transformation could be expressed by the transfer
matrix $L=L_j^k$ given by
\begin{equation}\label{trasfer matrix}
    L=\frac{1}{\sqrt{2}}\left(
        \begin{array}{cccc}
          1 & 0& 0 & 1 \\
           0 & 1 & i & 0\\
          0 & 1 & -i & 0 \\
           1 & 0 & 0 & -1\\
        \end{array}
      \right),\;\; L^{-1}=\frac{1}{\sqrt{2}}\left(
        \begin{array}{cccc}
          1 & 0& 0 & 1 \\
          0 & 1 & 1 & 0 \\
          0 & -i & i & 0 \\
          1 & 0 & 0 & -1 \\
        \end{array}
      \right).
\end{equation}
Thus,
\begin{equation}\label{coord tranf}
    y^\nu=L_\mu^\nu x^\mu,\; \;
    x^\nu=(L^{-1})_\mu^\nu y^\mu\,.
\end{equation}

The metric tensor $\eta$ in the Bondi tetrad is given by
\begin{equation}\label{NPmetric tensor}
    \tilde{\eta}=\left(
        \begin{array}{cccc}
          0 & 0 & 0 & 1 \\
          0 & 0 & -1 & 0 \\
          0 &  -1 & 0 & 0 \\
          1 & 0 & 0& 0 \\
        \end{array}
      \right)= \tilde{\eta}^{-1}.
\end{equation}

The bilinear symmetric scalar product of two  4-vectors  $\mathbf{a}=a^\mu \mathbf{n}_\mu$ and
$\mathbf{b}=b^\mu \mathbf{n}_\mu$ is given by
\[\mathbf{a}\cdot \mathbf{b}=\tilde{\eta} _{\mu\nu}a^\mu b^\nu. \]
This, for example, implies that
\begin{equation}\label{Zero Int NP}
    (\mathbf{a})^2=\mathbf{a}\cdot \mathbf{a}=0\;\;\Leftrightarrow\;\; a^0a^3=a^1a^2 \;\;\Leftrightarrow
    \;\; \frac{a^2}{a^0}=\frac{a^3}{a^1}\;\;\Leftrightarrow\;\; \frac{a^1}{a^0}=\frac{a^3}{a^2}\,.
\end{equation}
In case $a^0=0$, the last identities need to be reversed $\frac{a^0}{a^2}=\frac{a^1}{a^3},\;\frac{a^0}{a^1}=\frac{a^2}{a^3}.$
In these coordinates, the lowering of indices is denoted by $a_\mu=\tilde{\eta}_{\mu\nu} a^\nu$. For example,
\begin{equation}\label{LowerinIndexinNP}
   y_0=y^3,\;\;y_1=-y^2,\;\;y_2=-y^1,\;\;y_3=y^0.
\end{equation}

To understand the connection with the spinors of the representation $\tilde{\pi}$, which was based on the complex electromagnetic tensor  $\mathcal{F}$, defined by (\ref{complex tensor}) we  will present  this  tensor in Bondi tetrad. We will denote by $\widetilde{\mathcal{F}}$
the matrix of $\mathcal{F}$ in this representation. By the usual formula of basis transformation we get
\begin{equation}\label{FNP0}
    \widetilde{\mathcal{F}}^\beta_\alpha=L\mathcal{F}^\beta_\alpha L^{-1}=\frac{1}{2}\left(
                                         \begin{array}{cccc}
                                           F_3 & F_1-iF_2 & 0 & 0 \\
                                           F_1+iF_2 & -F_3 & 0 & 0 \\
                                           0 & 0 & F_3 &F_1-iF_2 \\
                                           0 & 0 & F_1+iF_2 & -F_3 \\
                                         \end{array}
                                       \right)=\sum_{j=1}^3 (\tilde{\rho}^j)^\alpha_\beta F_j.
\end{equation}
This show that the complex electromagnetic tensor become decomposable in Bondi tetrad. This tensor has two invariant subspaces $M_0=span_\mathbb{C}\{\mathbf{n}_0,\mathbf{n}_1\}$ and  $M_1=span_\mathbb{C}\{\mathbf{n}_2,\mathbf{n}_3\}.$

\begin{claim}\label{SpinRepPiTilde}
The representation $\tilde{\pi}$ of the Lorentz group on $M_c$ introduced in Definition \ref{pi-tildeDefn} acts as a spin half representation on each of subspaces  $M_0=span_\mathbb{C}\{\mathbf{n}_0,\mathbf{n}_1\}$ and  $M_1=span_\mathbb{C}\{\mathbf{n}_2,\mathbf{n}_3\}$
\end{claim}
From the decomposition (\ref{FNP0}) of the complex electromagnetic tensor  $\widetilde{\mathcal{F}}$, we see that in Bondi tetrad Majorana-Oppenheimer matrices $(\tilde{\rho}^j)^\alpha_\beta$ act on each $M_k$ as the the usual Pauli matrices $(\tilde{\rho}^j)^\alpha_\beta M_k=\sigma^jM_k$, with
\begin{equation}\label{PauliMatDef}
   \sigma^1=\left(
              \begin{array}{cc}
                0   & 1 \\
                1 &0 \\
              \end{array}
            \right),\;\; \sigma^2=\left(
              \begin{array}{cc}
                0   & -i \\
                i &0 \\
              \end{array}
            \right),\;\; \sigma^3=\left(
              \begin{array}{cc}
                1   & 0 \\
                0 &-1 \\
              \end{array}
            \right)\,.
\end{equation}
 Generators $\tilde{B}^j$ of boosts $\mathfrak{B}^j$ in direction of $x^j$ in representation $\tilde{\pi}$ were defined (Definition \ref{pi-tildeDefn}) by $\frac{1}{2}\rho^j$. So, on the subspaces $M_k$ they will act by $\tilde{\pi}(\tilde{B}^j)=\frac{1}{2}\sigma^k$. Similarly,  generators $\tilde{R}^j$ of rotation $\mathfrak{R}^j$ about the direction $x^j$ in representation $\tilde{\pi}$ were defined by $\frac{i}{2}\rho^j.$  So, on the subspaces $M_k$ they will act by $\tilde{\pi}(\tilde{R}^j)=\frac{i}{2}\sigma^k$. This identify the subspace $M_k$ with spinors and the representation $\tilde{\pi}$ with the spin half representation of the Lorentz group on the spinors. This proves the Claim.

The following tensor decomposition can help us to understand the connection between the representations $\tilde{\pi}$ and $\tilde{\pi}^*$. We define a tensor decomposition of a $4\times 4$ matrix as a
tensor product of $2\times 2$ matrices  by use of the binary representation of numbers. Each
of our indices $\mu=0,1,2,3$ can be considered as a pair of indices $(\mu_0,\mu_1)$ with value in $\mu_k\in\{0,1\}$ by
\[0\longmapsto 00,\;\;1\longmapsto 01,\;\;2\longmapsto 10,\;\;3\longmapsto 11\,.\]
The \textit{tensor decomposition} of a two tensor $D=D_{jk}$ is defined by
\begin{equation}\label{tesnor deconposition}
  D=a\otimes b, \;\;  D_{\mu\nu}=D_{(\mu_0,\mu_1)(\nu_0,\nu_1)}=a_{\mu_0\nu_0}\,b_{\mu_1\nu_1}\,.
\end{equation}
For example, the tensor $\tilde{\eta}$ can be decomposed as $\tilde{\eta}=\left(
                                                     \begin{array}{cc}
                                                       0   & 1 \\
                                                       -1 & 0 \\
                                                     \end{array}
                                                   \right)\otimes \left(
                                                     \begin{array}{cc}
                                                       0   & 1 \\
                                                       -1 & 0 \\
                                                     \end{array}
                                                   \right):=\tilde{\eta}_2\otimes \tilde{\eta}_2\,.$

The following properties of the tensor decomposition can be verified directly from the definition:
\[ a\otimes (b+c)=(a\otimes b)+( a\otimes c), \;\;k(a\otimes b)=(ka)\otimes b=a\otimes kb\]
and
\begin{equation}\label{tensor prop prod}
   ( a\otimes b)(c\otimes d)=ac \otimes bd,
\end{equation}
where $a,b,c,d$ are $2\times 2$ matrices and $k$ is a constant.

With this notation,the matrices $\rho^j$ in Bondi tetrad therefore become
\begin{equation}\label{KjinNP}
    (\widetilde{\rho}^j)^\mu_\nu=I_2\otimes \frac{1}{2}\sigma^j.
\end{equation}where $\sigma^j$ denote the  Pauli matrices, and $I_2$ is the  $2\times 2$ identity matrix.
In Bondi tetrad,
the tensor $(\mathcal{F}^*)^\alpha_\beta$ becomes
\[(\widetilde{\mathcal{F}}^*)^\alpha_\beta=L\mathcal{F}^* L^{-1}=\sum_j\bar{F}_j\sigma^*_j\otimes I_2\,.
\]
and
 \[ (\widetilde{\bar{\rho}}^k)^\mu_\nu=\sigma^*_j\otimes I_2\,.\]
This and (\ref{tensor prop prod}) explain why the representations $\tilde{\pi}$ and $\tilde{\pi}^*$
commute.

The connection  (\ref{FfromSAlf}) of the prepotential to the Faraday vector of the field is expressed by operators $\alpha^j:=\rho^j \mathcal{C}$. Like the Majorana-Oppenheimer matrices $\rho^j$, also these matrices satisfy the canonical anti-commutation relations (\ref{CAR}). In the  Bondi tetrad, see \cite{FGW}), the matrices $\alpha^j$ take the usual form  of the Dirac's $\alpha$-matrices
$\left(                \begin{array}{cc}
                  \sigma _j & 0 \\
                  0 & -\sigma _j \\
                \end{array}
              \right)$
   where $\sigma ^j$ are the Pauli matrices (see, for example \cite{Berestetskii}).

   Note that the matrices $\rho^j$, which define the representation $\tilde{\pi}$,
also satisfy the canonical anti-commutation relations (\ref{CAR}). However, the set $\rho_1,\rho_2,\rho_3$  cannot be completed by a forth anticommuting matrix, needed for the Dirac equation. Only after the conjugation they become Dirac's $\alpha$-matrices.

Parts of this paper were done in final undergraduate projects at Applied Physics Department at Jerusalem College of Technology by students S. Gwertzman, D. C. Gootvilig and M. Eliyahu under the supervision of the author. The author would like to thank  T. Scarr for editorial proof and E. Yudkin for help with the computer algebra  use of Mathematica.

\

\end{document}